\documentstyle[sprocl,epsfig,wrapfig]{article}

\def\qqbar{{\rm q}\overline{\rm q}}

\def\be{\begin{equation}}
\def\ee{\end{equation}}
\def\bea{\begin{eqnarray}}
\def\eea{\end{eqnarray}}

\def\ETJETs{(E_{\rm T}^{\rm jet})^2}
\def\etaa{\eta^{\rm jet1}}
\def\ETa{E_{\rm T}^{\rm jet1}}
\def\etab{\eta^{\rm jet2}}
\def\ETb{E_{\rm T}^{\rm jet2}}

\def\dspt{{\rm d}\sigma/{\rm d}p_{\rm T}^{\DST}}
\def\dseta{{\rm d}\sigma/{\rm d}\eta^{\DST}}
\def\DST{{\rm D}^{\ast}}

\begin{document}

\title{PHOTON STRUCTURE\footnote{invited talk given at DIS2000, Liverpool, UK, 25-30 April 2000}}

\author{STEFAN S\"OLDNER-REMBOLD}

\address{CERN, CH-1211 Geneva 23, Switzerland\\E-mail: 
stefan.soldner-rembold@cern.ch}


\maketitle\abstracts{The structure of the photon is probed 
in photon-photon interactions at LEP and in photon-proton
interactions at HERA. }

\section{Photon structure ?}
The photon is one of the fundamental gauge bosons of the Standard Model
without self-couplings and without intrinsic structure. 
It couples to any kind of charged particle, which allows it to
fluctuate directly into fermion-antifermion 
pairs and into bound states, vector mesons,
which have the same spin-parity ($J^{PC}=1^{--}$) as the photon. 
Photon-photon interactions therefore become possible through these quantum
fluctuations.

At low energies photon-photon interactions can be studied
e.g. in Delbr\"uck scattering (elastic scattering of photons in the electric
field of atoms) or in the elastic scattering process $\gamma\gamma\to\gamma
\gamma$ where existing experimental limits using lasers are still
18 orders of magnitude above QED predictions~\cite{bib-bernard}.
The knowledge of hadronic vacuum polarisations also still give the largest
systematic error in the QED predictions for g-2 and for the running
of the QED coupling constant~\cite{bib-gm2}.

The electron and positron beams at LEP and HERA can also be viewed
as a copious source of high energy quasi-real and virtual photons.
Interactions of photons are the hadronic processes with the
largest cross-section at these experiments. 
At these high energies fluctuation times are longer than
typical hadronic interaction time allowing the photon to
develop a `structure'.
The interactions 
of high energy photons can be described using structure functions and
parton distributions of the photon in analogy to the interactions of
real hadrons like the proton.

Why is it interesting to study photon structure ? Of course, it is simply
important to understand the high energy scattering of a fundamental
particle like the photon. Photon interactions are also sensitive to
the quantum structure of the theory and can therefore be used
to study QED and QCD. We can ask questions like: How
similar to the proton is the photon ? How are the quarks and gluons
distributed in the photon ? What happens if the photon becomes virtual ?
At future linear colliders it will be very important to understand
the large number of events from photon interactions, and last but not least
photon interactions can be an important probe to look for new
physics, for example in the production of Higgs
bosons at a 
photon linear collider ($\gamma\gamma\to\mbox{H}$)~\cite{bib-higgs}.

\section{Scales}
``The photon and the proton, who is probing whom ?'' asks Aharon 
Levy~\cite{bib-levy}. The same question can be asked for the
interactions of two photons. The answer is related to the physical scales 
in the process. The situation looks simple if only one physical
scale is large. 

At LEP we denote the virtuality of the ``probing'' photon with
$Q^2=-q^2$ (the negative squared four-momentum of the photon) 
and the virtuality of the ``probed'' photon with $P^2=-p^2\approx 0$. Just like
for the proton, the deep-inelastic scattering cross-section is than
written as
 \begin{equation}
 \frac{{\rm d}^2\sigma_{\rm e\gamma\rightarrow {\rm e+hadrons}}}{{\rm d}
x{\rm d}Q^2}
 =\frac{2\pi\alpha^2}{x\,Q^{4}}
  \left[ \left( 1+(1-y)^2\right) F_2^{\gamma}(x,Q^2) - y^{2}
F_{\rm L}^{\gamma}(x,Q^2)\right],
\label{eq-eq1}
 \end{equation}
where $\alpha$ is the fine structure constant, $x$ and $y$ are
are the usual dimensionless variables of deep-inelastic scattering and
$W^2=(q+p)^2$ is the squared invariant mass of the hadronic final state.
The scaling variable $x$ is given by 
\begin{wrapfigure}{r}{0.5\textwidth}
\epsfig{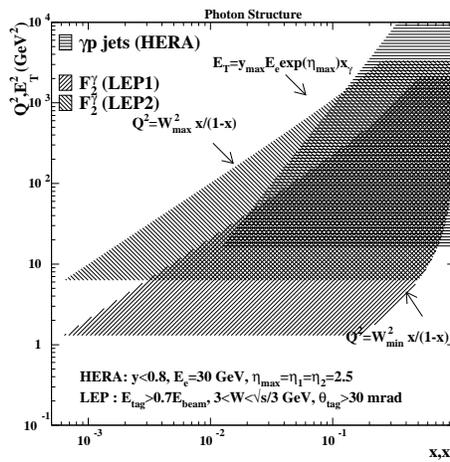}
\caption{A comparison of the $(Q^2,x)$ planes covered by LEP
with the $(E_T^2,x_{\gamma})$ plane covered by HERA studying
jet production.}
\label{fig-kin}
\end{wrapfigure}
\begin{equation}
x=\frac{Q^2}{Q^2+W^2+P^2}.
\end{equation}
The term proportional to $F_{\rm L}^{\gamma}(x,Q^2)$ is small
and is therefore usually neglected. The structure function 
$F_2^{\gamma}(x,Q^2)$ can be identified with the sum over the parton
densities of the photon weighted by the square of the parton's charge.
As a consequence deep-inelastic scattering mainly probes the
quark structure of the photon, gluons only enter through scaling violations.

Another process in which photon structure can be studied is
the production of (di-)jets in photon-proton (HERA) or photon-photon (LEP)
interactions. The interacting photons are now almost real and
the largest physical scale is the transverse energy of
the jets. The variable $x_{\gamma}$ which is related
to the fraction of the photon's momentum participating in
the hard interactions can be reconstructed 
from the pseudorapidities $\eta^{\rm jet}$ 
and transverse energies $E_{\rm T}^{\rm jet}$ of the jets:
\begin{equation}
x_{\gamma}=\frac{\ETa e^{-\etaa}+\ETb e^{-\etab}}{2yE_{\rm e}},
\end{equation}
where $yE_{\rm e}$ is the energy taken by the photon. 
In leading order, $x_{\gamma}$ 
is equivalent to $x$ and we can relate the parton 
distributions probed at LEP and HERA by $q_{\rm LEP}(Q^2,x)\approx
q_{\rm HERA}(\ETJETs,x_{\gamma})$. In jet production gluon induced
processes dominate the cross-section in most kinematic regions, i.e.
different from deep-inelastic electron-photon scattering the results
are directly sensitive to the gluon distribution in the photon.

In Fig.~\ref{fig-kin} the kinematic
range accessible at LEP and HERA is compared. At HERA accessing 
the low $x$ parton densities of the photon requires the reconstruction
of jets at low $E_{\rm T}^{\rm jet}$ and large $\eta^{\rm jet}$. This
is experimentally difficult. In addition, additional soft or hard
interactions of the photon's and the proton's remnant can take
place which need to be disentangled from the primary hard scattering process. 

\section{How proton-like is the photon ?}
The question should really be ``How $\rho$-like is the photon ?'', but
since we know much more about proton structure~\cite{bib-erdmann}
than $\rho$ structure and since we study photon-proton interactions,
we take the proton as a generic hadron for comparisons.

In a simple picture we can split the structure function $F_2^{\gamma}$
into two parts, a Vector Mesons Dominance (VMD) part 
where the photon has fluctuated into
a bound ``hadron-like'' state and a ``point-like'' part where the photon
couples directly to a quark-antiquark pair. The original interest
in the photon structure function was driven by the point-like part which
can be calculated in the Quark-Parton-Model (QPM),
\begin{eqnarray}
\lefteqn{F_2^{\gamma,\rm QPM}(x,Q^2)=}\nonumber\\ & &
\frac{N_{\rm c}\alpha}{\pi}\sum_{\rm q} e_{\rm q}^4 x\left[ (x^2+(1-x)^2)
\left( \ln\frac{Q^2}{m_{\rm q}^2}-\ln\frac{x}{1-x}\right) -1+8x(1-x)\right]
\end{eqnarray}
which is equivalent to the QED structure function. The structure function
$F_2^{\gamma,\rm QPM}(x,Q^2)$ rises linearly with 
$\ln\frac{Q^2}{m_{\rm q}^2}$. It was first shown by 
Witten~\cite{bib-witten}
that this linear rise is still expected if QCD is turned on, however
the parton mass $m_{\rm q}$ is replaced by the QCD scale $\Lambda_{\rm QCD}$.
For asymptotically large $Q^2$ the structure function is fully calculable,
including the normalisation. This is called the asymptotic prediction.

\begin{figure}[htbp]
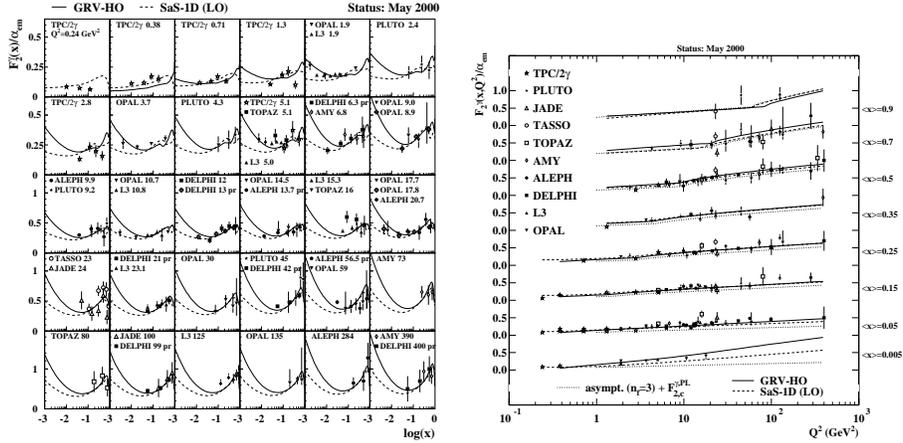

\begin{tabular}{cc}
\epsfig{file=f2all_bw.epsi,width=0.48\textwidth}
 &
\epsfig{file=f2q2_bw.epsi,width=0.48\textwidth}
\end{tabular}
\caption{ 
Measurements of the photon structure function $F_2^{\gamma}$
in bins of $x$ and $Q^2$. }
\end{figure}

In the real world - far away from asymptotia - we have to take into
account the non-perturbative ``VMD-like'' part for which some ansatz
at small $Q^2$ is chosen and subsequently evolved with the inhomogeneous
DGLAP equations. ``Inhomogeneous'' refers to the additional term which
enters the evolution already in leading order, the perturbative splitting
$\gamma\to\qqbar$. As a consequence the photon structure function
exhibits positive scaling violations (it rises with $Q^2$) 
at all values of $x$, not only at low $x$ as for the proton. 
This is the most striking difference between the photon and the proton.

\section{How many gluons in a photon ?}
At low $x$, approximately in the range $x<0.1$, 
we expect that the structure function of the photon rises towards low $x$
and with increasing $Q^2$ driven by
the same QCD evolution as the proton structure function.
Ideal processes to study the gluon in the photon are the production
of (di-)jets and the production of heavy quarks in photon induced
interactions.

\subsection{Charm in or from the photon ?}

With charm a new scale enters, the mass $m_{\rm c}$ of the charm
quark. The main LO contributions
to open charm production in photon-photon interactions at LEP 
are the direct and the single resolved process.
The single resolved photon-photon process is analogous to
the direct photon-proton process where the gluon is taken
from the proton and not from one of the photons. In both cases
we can probe the gluon content, either of the photon or of the proton.
Double resolved processes are negligible for open charm production
in photon-photon interactions, but they are important in photoproduction.
 
The NLO calculations of open charm production
are either done in the so-called massive or in the so-called
massless scheme.
\begin{figure}
\begin{tabular}{ccc}
\epsfig{file=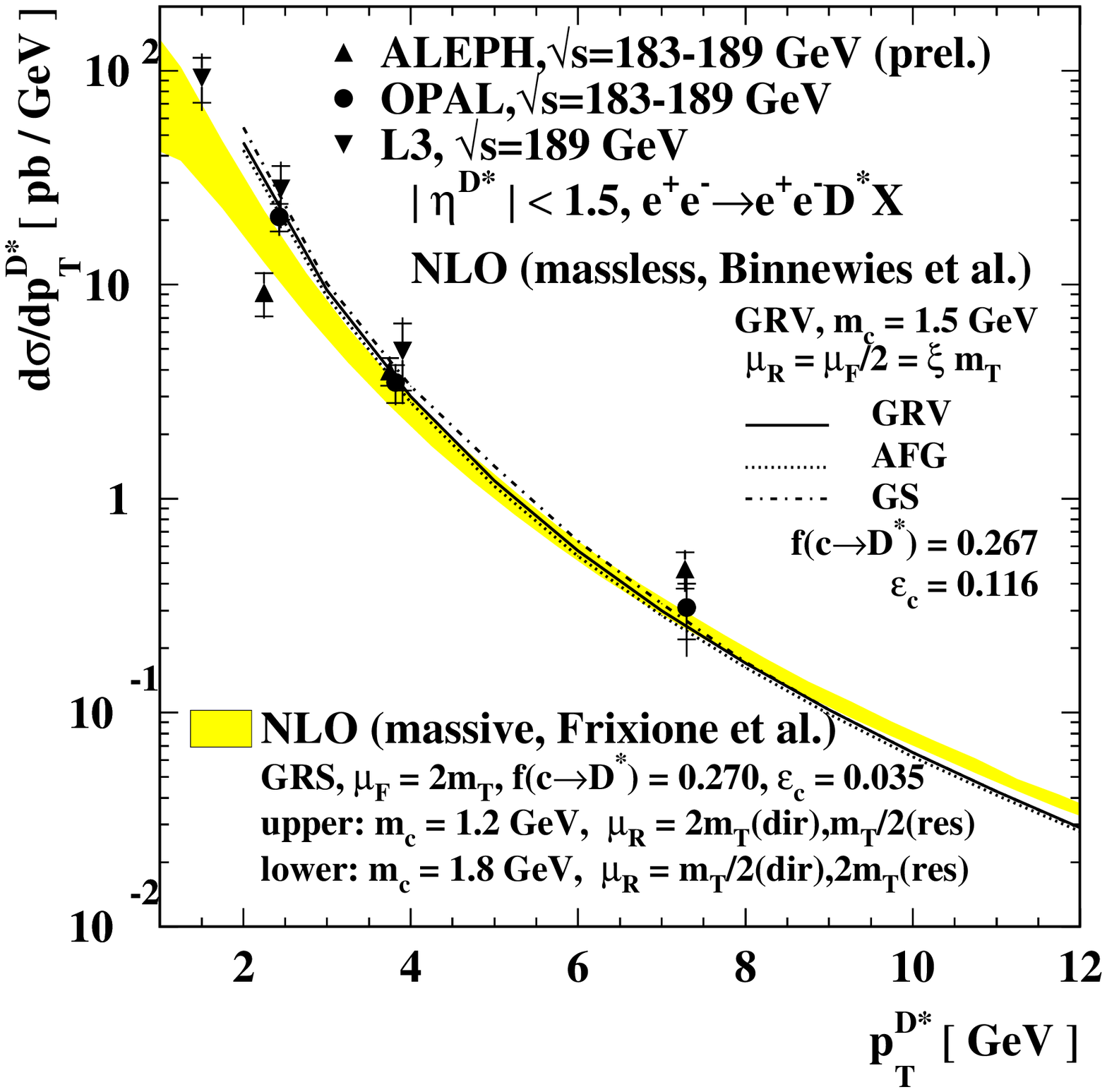,width=0.4\textwidth} &
\epsfig{file=fig3iacol.epsi,width=0.25\textwidth} &
\epsfig{file=fig3ibcol.epsi,width=0.25\textwidth} 
\end{tabular}
\caption{
a) Differential cross-section $\dspt$
measured in photon-photon interactions by L3 and OPAL.
b) Differential cross-section $\dseta$
measured by ZEUS in photoproduction
compared to NLO calculations 
by Kniehl et~al.~\protect\cite{bib-Kniehl2} (left) and by 
Cacciari et~al.~\protect\cite{bib-cacc} (right).}
\label{fig-sigpt}
\end{figure}
In the massive scheme the mass $m_{\rm c}$ of the charm quark sets
the scale for the perturbative QCD calculation. The cross-section
is factorized into the matrix elements for the production
of heavy quarks and the parton densities for light quarks (uds) and
gluons. This `massive' approach is expected to be valid
if the transverse momenta $p_{\rm T}$ of the charm quarks are of the
same order as the charm mass, $p_{\rm T} \approx m_{\rm c}$.
In the `massless' scheme, charm is considered as one of
the active flavours in the parton distributions like u,d,s.
This scheme is expected to be valid for $p_{\rm T} >> m_{\rm c}$.

Open charm is usually tagged by measuring the production
of D$^*$ mesons.
In Fig.~\ref{fig-sigpt}a, the differential cross-section $\dspt$
measured in photon-photon interactions by L3 and OPAL~\cite{bib-dlep}
is compared to the NLO calculation by Frixione et 
al.~\cite{bib-Frixione} using the massive approach and to the NLO calculation
by Kniehl et~al.~\cite{bib-Kniehl1} using the massless approach.
The massless calculation is in better agreement with the
data than the massive calculation.
In Fig.~\ref{fig-sigpt}b, the differential cross-section $\dseta$
measured by ZEUS in photoproduction
is compared to two different massless calculations,
by Kniehl et~al.~\cite{bib-Kniehl2} and by Cacciari et~al.~\cite{bib-cacc}.
The NLO calculations tend to underestimate the cross-section
in the forward direction. Whereas in photon-photon scattering
the massless cross-section seems to be nearly independent
of the parton densities used, there seems to be more sensitivity
to the choice of parametrisation in photoproduction.

\subsection{Jet production }
Another way to introduce a new hard scale in the process is
to study jet production at large transverse 
momenta~\cite{bib-jets,bib-wing,bib-max}.
Different groups have followed different philosophies for extracting
information about the parton (mainly gluon) content of the photon from
jet cross-sections. In the first approach hadronic jet cross-sections
are measured and compared to NLO calculations which use
different parametrisations of the photon's parton densities
as input. In the second approach
LO parton densities are extracted from the measurements.

\begin{figure}[htbp]
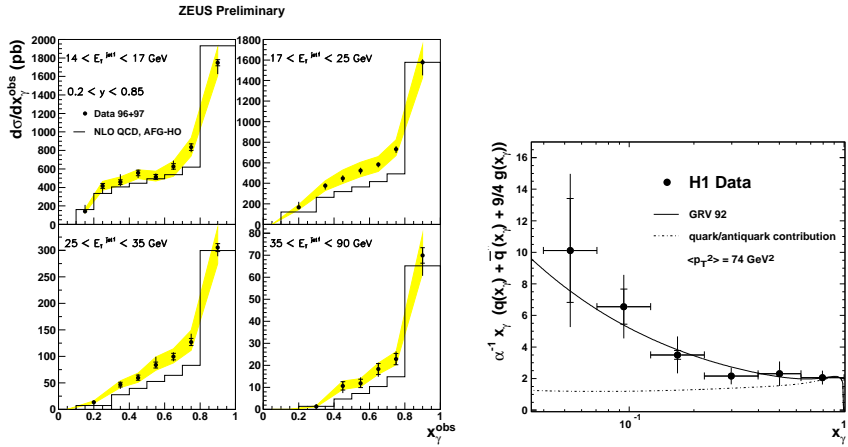

\begin{tabular}{ccc}
\epsfig{file=xgamma.epsi,width=0.5\textwidth} &
\epsfig{file=d00-035f4.epsi,width=0.4\textwidth} &
\end{tabular}
\caption{a) Differential cross-section 
${\rm d}\sigma/{\rm d}x_{\gamma}^{\rm obs}$ for different
$E_{\rm T}^{\rm jet}$ bins using jets in the range
$-1<\eta_{\rm jet}<2$ and $Q^2<1$~GeV$^2$ (ZEUS).
b) Effective parton density of the photon
for average jet transverse momenta $\langle
p_{\rm T}^2\rangle=74$~GeV$^2$ (H1).}
\label{fig-jets}
\end{figure}

In di-jet production the variable $x_{\gamma}$ can
be reconstructed from the momenta of the two jets.
Fig.~\ref{fig-jets}a shows a ZEUS measurement of 
the differential cross-section 
${\rm d}\sigma/{\rm d}x_{\gamma}^{\rm obs}$ for different
$E_{\rm T}^{\rm jet}$ bins using jets in the range
$-1<\eta_{\rm jet}<2$ and $Q^2<1$~GeV$^2$~\cite{bib-wing}. 
The NLO calculations lie systematically too low which could indicate
the need for more gluons in the parametrisations of the
parton densities. NLO calculations are performed at
the parton level and contain no hadronisation effects
and also no underlying event. In addition, scale uncertainties
have to be taken into account. For the high $E_{\rm T}^{\rm jet}$ 
region considered here, these effects are expected
to be small enough so that the discrepancy between data and
the NLO calculations can be attributed to inadequacies 
of the parametrisations of the parton densities.

The second approach to extract effective parton densities
is shown in Fig.~\ref{fig-jets}b. The effective parton
densities are extracted from the di-jet data assuming
a similar angular distribution for all resolved 
processes~\cite{bib-max}.
The effective parton density of the photon is given by
\begin{equation}
\tilde{q}_{\gamma}(x_{\gamma},p_{\rm T}^2)
 \equiv \sum_{n_f}\left(
 q_{\gamma}(x_{\gamma},p_{\rm T}^2)
+\overline{q}_{\gamma}(x_{\gamma},p_{\rm T}^2)
\right)+\frac{9}{4}g_{\gamma}(x_{\gamma},p_{\rm T}^2).
\end{equation}
The LO quark density $q_{\gamma}(x_{\gamma},p_{\rm T}^2)
+\overline{q}_{\gamma}(x_{\gamma},p_{\rm T}^2)$ is reasonable well constrained
by e$^+$e$^-$ data in this kinematic range and its contribution 
- shown as 
\begin{wrapfigure}[20]{r}{0.5\textwidth}
\epsfig{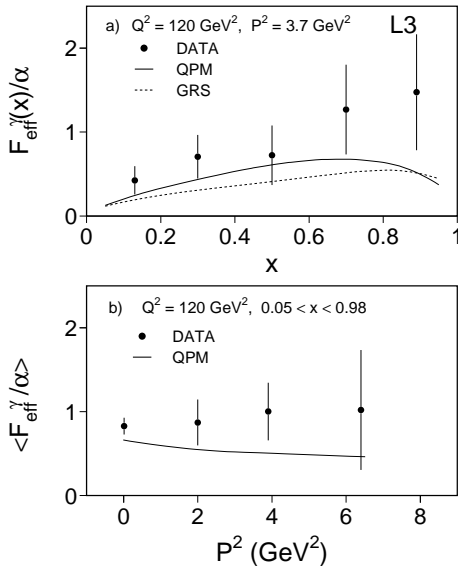}
\caption{
Effective structure function measured by L3 as function of $x$
and $P^2$.}
\label{fig-f2eff}
\end{wrapfigure}
dashed line in Fig.~\ref{fig-jets} - is small. 
If the quark distribution is subtracted, a clear rise
of the gluon distribution towards low $x$ can be observed.

\section{Parton densities of the virtual photon}
Until now we have studied the structure of
(quasi-)real photons, i.e. $P^2\approx 0$~(LEP) or $Q^2\approx 0$~(HERA). 
In e$^+$e$^-$ collisions the effective structure function
of virtual photons can be measured if $Q^2>>P^2>>\Lambda_{\rm QCD}$.
This was first done by PLUTO~\cite{bib-pluto}.
For real photons only
the cross-sections $\sigma_{\rm LT}$ and $\sigma_{\rm TT}$
contribute, where the indices refer to the longitudinal and transverse
helicity states of the probe and target photon, respectively,
i.e. $F_2^{\gamma}\simeq \sigma_{\rm LT}+\sigma_{\rm TT}$.
For $P^2>>0$ other helicity states have to be taken into
account, leading to the definition of the effective
structure function
$F_{\rm eff}^{\gamma}\simeq \sigma_{\rm LT}+\sigma_{\rm TT}
+\sigma_{\rm TL}+\sigma_{\rm LL}$ (interference terms are neglected
here). This effective structure function has been measured
by L3 and is shown in Fig.~\ref{fig-f2eff}.
We expect the non-perturbative part of the parton densities (VMD) at low
$x$ to decrease with increasing virtuality of the photon. 
Compared to the data as a function of
$P^2$ in Fig.~\ref{fig-f2eff}b, the QPM prediction therefore
fails to describe the point at $P^2=0$. The shape of the $P^2$
dependence is consistent with the simple QPM ansatz but the
errors are still large. Much more precise data is to be expected
from LEP on the structure of the virtual photons in the next years.

\section{$\gamma^*\gamma^*$ scattering}
For studying the effective structure function of virtual photons, we
assumed that $Q^2>>P^2$. In the special case where both
photons have large and approximately equal virtualities,
$Q^2\approx P^2$ (or better $Q_1^2\approx Q_2^2$), the structure
function formalism can no longer be applied. 

\begin{figure}[htbp]
\begin{tabular}{ccc}
\epsfig{file=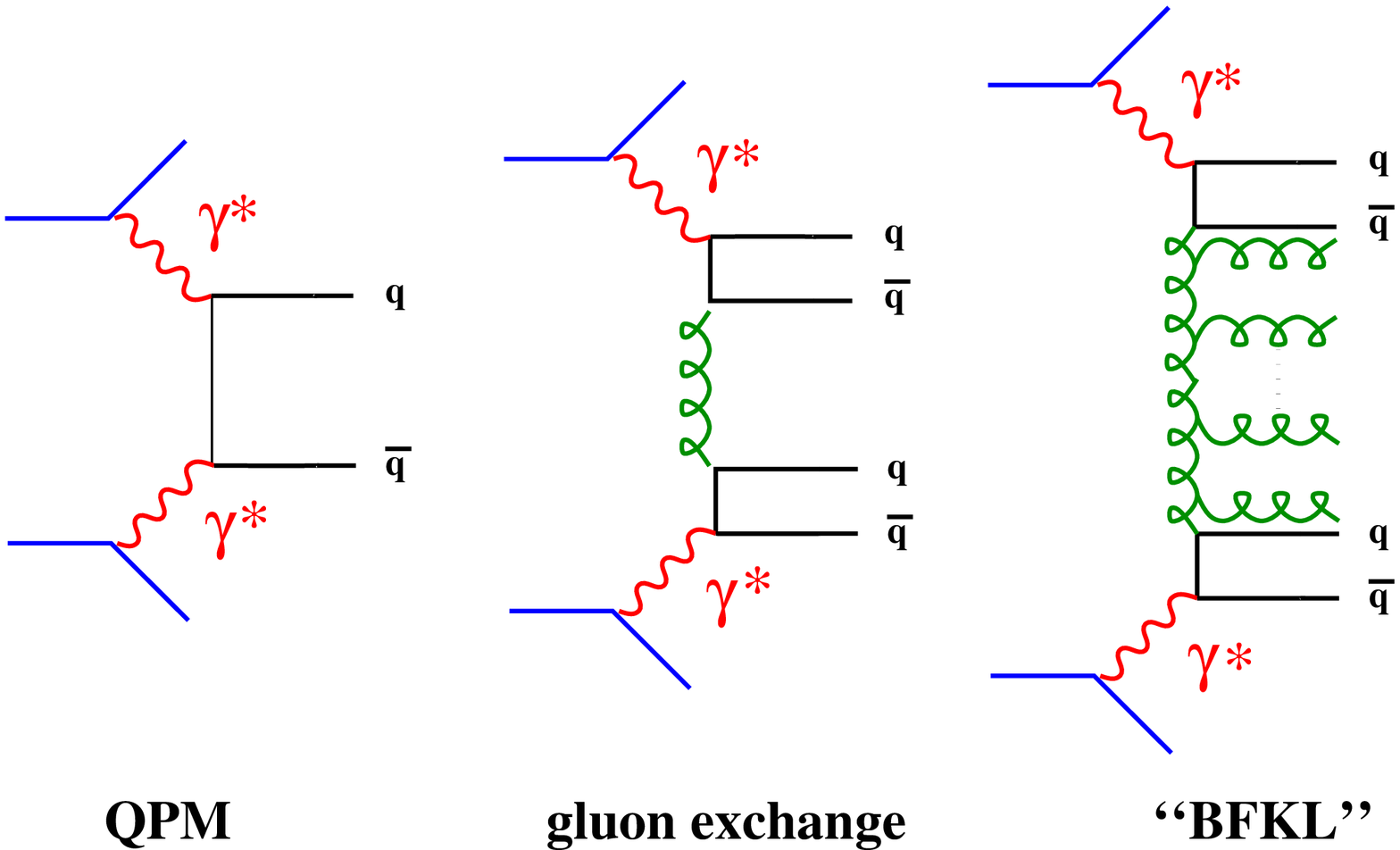,width=0.5\textwidth} &
\epsfig{file=bfklpf3c.epsi,width=0.4\textwidth} &
\end{tabular}
\caption{a) Different diagrams which are expected to
contribute to $\gamma^*\gamma^*$ scattering: QPM (Quark Parton Model),
single gluon exchange and the `BFKL' process.
b) the hadronic $\gamma^*\gamma^*$ cross-section measured by L3
as a function of $Y$ compared to the calculations of~\protect\cite{bib-kim}.}
\label{fig-bfkl}
\end{figure}

For sufficiently large virtualities this process has been
called the `optimal test' of the prediction of the BFKL 
formalism~\cite{bib-bfklcharm}. The condition $Q_1^2\approx Q_2^2$
ensures that DGLAP evolution is suppressed.
The application of the BFKL formalism to $\gamma^*\gamma^*$ scattering
has been considered by~\cite{bib-bfkl,bib-kim,bib-brodsky}.
A sketch of the main diagrams is shown in Fig.~\ref{fig-bfkl}a.

In the BFKL formalism there is a problem at LO in setting the two mass
scales on which the cross-section depends: the mass at which the 
strong coupling $\alpha_s$ is evaluated and the mass which provides 
the scale for the high energy logarithms. The result is very sensitive to 
these parameters~\cite{bib-brodsky}.
An additional uncertainty is due to the correct treatment of the production
of massive charm quarks~\cite{bib-bfklcharm}.

From the LO BFKL prediction an approximately exponential increase
of the hadronic cross-section $\sigma(\gamma^*\gamma^*)$ is expected
as a function of the variable 
\begin{equation}Y\approx\ln\frac{W^2}{\sqrt{Q_1^2Q_2^2}}.\end{equation}
The LO BFKL cross-section is significantly too high compared to the 
L3 data shown in Fig.~\ref{fig-bfkl}b where the data are compared to
the calculations of~\cite{bib-kim}. A prediction which includes
an estimate of the NLO effects is also shown. The NLO curves
are much closer to the data.
It is also interesting to note that 
the LO Monte Carlo PHOJET is consistent with the data within the large
experimental errors, apart from the very high $Y$ region where deviations
are seen.
Both, the theoretical and experimental uncertainties are still large
and require further study before more quantitative conclusions can be 
drawn~\cite{bib-bfklhere}.

\section{Conclusions}
The amount of interesting data from LEP and HERA on
the structure of real and virtual photons has increased
so much in the last years that only a few highlights
can be mentioned in such a brief write-up. 
I have not discussed the interesting H1 measurements
of the effective parton densities of the virtual photons, the
ZEUS measurements of the suppression of the resolved component
with increased photon virtuality, the OPAL measurement of
the charm structure function of the photon, and many other topics.

It becomes more and more clear that we need a common quantitative
understanding of the LEP and HERA data on photon structure,
especially in the form of new parton densities which make use of the new
precise data from LEP and HERA. To accomplish this, we also
still need a better understanding of the non-perturbative effects
(hadronisation, underlying event) which still lead to systematic
limitations in the interpretation of the data.

\section*{Acknowledgments}
I would like to thank John Dainton and his colleagues from Liverpool
for organising this interesting and exciting conference.


\section*{References}
\begin{thebibliography}{99}
\bibitem{bib-bernard} 
D.~Bernard, Nucl.~Phys.~B (Proc.~Suppl.) 82 (2000) 439.
\bibitem{bib-gm2}
M.~Davier, A.~H\"ocker, Phys.~Lett.~B435 (1998) 427.
\bibitem{bib-higgs} 
M.~Melles, Nucl.~Phys.~B (Proc.~Suppl.) 82 (2000) 379;
G.~Jikia, S.~S\"oldner-Rembold, Nucl.~Phys.~B (Proc.~Suppl.) 82 (2000) 373.
\bibitem{bib-levy}
A.~Levy, Proc. of 1999 KEK-Tanashi Symposium, hep-ph/0002015.  
\bibitem{bib-erdmann}
M.~Erdmann, these proceedings.
\bibitem{bib-witten}
E.~Witten, Nucl.~Phys.~B120 (1977) 189.
\bibitem{bib-dzeus}
ZEUS Collaboration, J.~Breitweg et al., Eur.~Phys.~J.~C6 (1999) 67.
\bibitem{bib-dlep}
L3 Collaboration, M.~Acciarri et~al., Phys.~Lett.~B467 (1999) 137;
OPAL Collaboration, G.~Abbiendi et~al., hep-ex/9911030.
\bibitem{bib-Frixione}
S.~Frixione et al, hep-ph/9908483.
\bibitem{bib-Kniehl1}
J.~Binnewies, B.A.~Kniehl and G.~Kramer, Phys.~Rev.~D58 (1998) 014014;  
Phys.~Rev.~D53 (1996) 6110.
\bibitem{bib-Kniehl2}
B.~A.~Kniehl et al., Z.~Phys.~C76 (1997) 689.
\bibitem{bib-cacc}
M.~Cacciari et al.,  Phys.~Rev.~D55 (1997) 2736; ibid 7134.
\bibitem{bib-jets}
E.~Heaphy, B.~Surrow, A.~Valkarova, these proceedings.
\bibitem{bib-wing}
M.~Wing, these proceedings, hep-ex/0007011.  
\bibitem{bib-max}
S.~Maxfield, these proceedings;
H1 Collaboration, C.~Adloff et al., Phys.~Lett.~B483 (2000) 36.
\bibitem{bib-pluto}
PLUTO Collaboration, C.~Berger et al., Phys.~Lett.~B142 (1984) 119.
\bibitem{bib-bfklcharm}
J. Bartels, C. Ewerz, R. Staritzbichler, hep-ph/0004029. 
\bibitem{bib-bfkl}
F. Hautmann: Proceedings of the XXVIII International 
Conference on High Energy Physics ICHEP96 (Warsaw, July 1996), 
eds. Z. Ajduk and A.K. Wroblewski, World Scientific, p.705;
J. Bartels, A. De Roeck and H. Lotter: Phys.~Lett. {B389}
(1996) 742;
M. Boonekamp, A. De Roeck, C. Royon and S. Wallon: hep-ph/9812523;
J. Kwiecinski and L. Motyka: Phys.~Lett.~B462 (1999) 203; 
N.N. Nikolaev, J. Speth, V.R. Zoller, hep-ph/0001120; 
\bibitem{bib-kim}
V.T. Kim, L.N. Lipatov, G. Pivovarov hep-ph/9911242, hep-ph/9911228; 
\bibitem{bib-brodsky}
S.J. Brodsky, F. Hautmann and D.E.Soper: Phys.Rev.Lett.
{78} (1997) 803; Erratum ibid {79} (1997) 3544;
S.J. Brodsky, F. Hautmann and D.E. Soper: Phys.Rev. {D56}
(1997) 6957.
\bibitem{bib-bfklhere}
A. Donnachie, C.-H. Lin, L. Lipatov, E. Naftali, these proceedings.
\end{thebibliography}
\end{document}